\documentclass[a4paper,twoside,12pt,notitlepage]{article} 
\usepackage[english]{babel}
\usepackage{graphicx}
\usepackage[T1]{fontenc}
\usepackage[cp1250]{inputenc}
\usepackage{appendix}
\usepackage{verbatim}
\usepackage{amsmath}
\usepackage{amsfonts}
\usepackage{ragged2e}
\allowdisplaybreaks

\newcommand\al{\alpha}
\newcommand\be{\beta}
\newcommand\ga{\gamma}

\newcommand\de{\delta}

\newcommand\eps{\epsilon}

\newcommand\la{\lambda}

\newcommand\si{\sigma}

\newcommand\cD{{\mathcal D}}

\newcommand\cL{{\mathcal L}}

\newcommand\cO{{\mathcal O}}

\newcommand\cS{{\mathcal S}}

\newcommand\AAA{{\mathbb A}}
\newcommand\one{{\bf{1}}}

\newcommand\OOO{{\mathbb O}}

\newcommand\rd{{\rm d}}
\newcommand\re{{\rm e}}
\newcommand\ri{{\rm i}}

\newcommand\beq{\begin{equation}}
\newcommand\eeq{\end{equation}}
\newcommand\beaX{\begin{eqnarray*}}
\newcommand\eeaX{\end{eqnarray*}}
\newcommand\bea{\begin{gather}\begin{aligned}}
\newcommand\eea{\end{aligned}\end{gather}}
\newcommand\btp{\begin{tikzpicture}}
\newcommand\etp{\end{tikzpicture}}
\newcommand\nn{\notag}

\newcommand\rarr{\rightarrow}

\newcommand\pd{\partial}

\newcommand\Tr{\text{Tr}}

\newcommand\pslash{p \hspace{-1.0ex}\slash}

\newcommand\Pexp{\mathcal{P}\hspace{-.4ex}\exp}

\begin{document}
\begin{center}
{\bf \Large Manifestly gauge-covariant representation of scalar and fermion propagators}\\[1.5cm]

{\bf Adam Latosi{\'n}ski} \\
{{\it Max-Planck-Institut f\"ur Gravitationsphysik
(Albert-Einstein-Institut)\\
Am M\"uhlenberg 1, D-14476 Potsdam, Germany}}
\end{center}

\vspace{2cm}

{\footnotesize

A new way to write the massive scalar and fermion propagators on a background of a weak gauge field is presented. They are written in a form that is manifestly gauge-covariant up to several additional terms that can be written as boundary terms in momentum space. These additional terms violate Ward-Takahashi identities and need to be renormalized by appropriate counterterms if the complete theory is to be gauge-covariant. This form makes it possible to calculate many amplitudes in a manifestly gauge-covariant way (at the same time reducing the number of Feynman diagrams). It also allows to express some counterterms in a way independent of the regularization scheme and provides an easy way to derive the anomalous term affecting the chiral current conservation.
}
\newpage

\section{Introduction}

In a quantum field theory, when one wants to calculate a specific amplitude in the form of a formal series, one often uses Feynman diagrams \cite{Feynman}. This is doubtlessly an amazing method which allows to swiftly write a number of expressions contributing to the desired amplitude. However it is not without weaknesses. Because it is based on a division of the lagrangian into the free-field lagrangian, the interaction lagrangian and the counterterms, the individual expressions often do not have the symmetries of the theory if the parts of the lagrangian do not posses them on their own. One such symmetry that is never manifestly preserved is the gauge symmetry, since the initially gauge-invariant terms in the lagrangian are broken up into pieces, and different pieces are treated differently. Moreover, if we consider the gauge field to be not only the classic, background field, but a dynamic quantum field, we need to add the gauge fixing term to the lagrangian so that we could construct a propagator for the gauge field. Eventually the individual expressions are not gauge-covariant, and even when summed up the gauge-covariance of the final result is often not visible at once, and is only recovered because of cancellations between many terms. Sometimes even the Green functions themselves aren't gauge invariant, and only the physical observables, like cross-sections and scatering amplitudes, are. \\
\indent The alternative approach by Schwinger \cite{Schwinger} and Tomonaga \cite{Tomonaga} is gauge covariant, but at the same time it is difficult to calculate with and opaque. \\
\indent While the development of the numerical method in last decades and rising power of the computers made it possible to calculate many diagrams with relative ease, at the same time, by receiving only the final result we lose the insight on how it came to be. Algorithms that perform the calculations rarely even give the result in the form of a gauge-covariant formula, producing instead only numerical values that one needs a lot of experience to comprehend easily. I believe that to preserve the understanding of the gauge theory one should look for analytical expression if possible, and be more aware of how certain results are produced. However, the method I'm going to present is also algorithmic and can be put on a computer if you wish so. \\
\indent I am going to present a way of obtaining the formulae for scalar and fermion propagators that are gauge invariant and can be used as building blocks in Feynman diagrams. It bases on the background field method \cite{DeWitt, Abbott, AbbotGrisaruSchaefer} used to calculate the effective action. That method already guarantees that the final result is gauge-covariant, but in this work I want to formulate such Feynman rules that not only the final result, but also intermediate steps are given by gauge-covariant expressions. It is done by expressing the propagators of the particles on the background of a gauge field using path-ordered exponentials and a quasi-local formal series made of exclusively gauge-covariant quantities. One may be familiar with the following approximation for the quark propagator in a background of gluon fields:
\beq \langle q(x_1) \bar{q}(x_2)\rangle \approx \Pexp\big(-\ri g\int^{x_1}_{x_2} G^a_\mu(x) T^a \rd\hspace{-.2ex}x^\mu\big) \langle q(x_1) \bar{q}(x_2)\rangle_0 \eeq
where $\Pexp$ denotes a path-ordered exponential (along some path connecting points $x_1$ and $x_2$) and subscript 0 refers to free quark fields. Typically, one needs to average over all possible paths, the concept that is used in the lattice gauge theory introduced by Wilson \cite{Wilson}. The idea I follow in this work is to focus one path - the straight line - and find the corrections to the formula above in the form of a series constructed with gauge-covariant operators. \\ 
\indent First of all, the corrections I find exhibit nontrivial spinor structure, the true propagator cannot be expressed simply by multiplying the free propagator with some function of the gauge field. Notably, amongst other terms in the propagator, I find one proportional to $\ga_5$, which can be used to obtain the chiral anomaly (ABJ-anomaly \cite{Adler, BellJackiw}) in a new, transparent way, without the need of any regularization. The full formula for the propagator, in the form of a series, can also be used in Feynman diagrams, making them to produce only gauge-invariant expressions and sparing us the need of calculating terms that would cancel each other in the final result anyway. \\
\indent In this work, I construct the formulae for the scalar and spinor propagators. For the full set of rules, we would also need the rules for the lines of gauge bosons (which in the backgound field method never appear as external lines, but can appear inside the diagrams). However, this case is much more complicated, especially if we want to consider a general gauge (which may also include ghosts) or spontanous symmetry breaking like in the SM. Since even with only scalars and fermions we can see some interesting features of the method I present, and because of the limited space I have, I decided to leave the vector boson case out of this paper, and I plan to publish it in a following work.

\section{Presentation of the method, scalar case}
Let us consider a theory of a scalar field minimally coupled to an external gauge field:
\beq \cL = D^\mu\phi^\dag D_\mu\phi - m^2 \phi^\dag\phi \label{LagrangianScalars} \eeq
where
\beq D_\mu\phi = \pd_\mu\phi - \AAA_\mu\phi \nn \eeq
\beq \AAA_\mu = -\ri g A_\mu^a T^a = -\AAA_\mu^\dag \label {CovariantDerivative} \eeq
For now, we are going to consider the field $\AAA_\mu$ to be a background field, so we do not include its kinetic energy in the lagrangian, and it will not appear as a propagating field inside the diagrams. This lagrangian is invariant under a local redefinition of fields with $U(x) =\exp(-\ri g \theta^a(x) T^a)$:
\beq \phi(x) = U(x)\phi'(x),\qquad \AAA_\mu(x) = U(x)\AAA'_\mu(x)U(x)^{-1} + \pd_\mu U(x)U(x)^{-1} \label{GaugeTransformation} \eeq 

\noindent In the path integral formulation of the quantum theory, we have
\beq Z_\phi[J,J^\dag,A] = \re^{\ri W[J,J^\dag,A]} = \int \cD\phi\, \re^{\ri \int\rd^4x \big(\cL[A] + J^\dag\phi + \phi^\dag J\big)} \label{GeneratingFunctionalScalar} \eeq 

\noindent Let us consider the 2-point Green function
\beq G_\phi(x_1,x_2)[A] = \langle \phi(x_1)\phi^\dag(x_2) \rangle= -\ri\frac{\de^2W_\phi[J,J^\dag,A]}{\de J^\dag(x_1)\de J(x_2)}\bigg|_{J=0} \label{2PGFScalar} \eeq
This function should follow the appropriate gauge covariance rule
\beq G(x_1,x_2)[A] = U(x_1)G(x_1,x_2)[A']U(x_2)^{-1} \eeq
which gives us the following Ward-Takahashi identity: 
\begin{align} & \frac{\pd}{\pd x^\mu}\frac{\de G_\phi(x_1,x_2)[A]}{\de A^a_\mu(x)} - g f^{abc} A_\mu^b(x) \frac{\de G_\phi(x_1,x_2)[A]}{\de A^c_\mu(x)} = \nn\\
&\qquad = -\ri g \bigg(G_\phi(x_1,x_2)[A]T^a\de(x_2-x)-\de(x_1-x)T^a G_\phi(x_1,x_2)[A]\bigg) \label{WTIdentity} \end{align}

The issue we are going to address is the fact that when we calculate Green function $G_2(x_1,x_2)[A]$ using perturbative expansion, the satisfaction of eq. (\ref{WTIdentity}), which comes from the gauge-covariance, is completely invisible. The formulae we get are not-gauge-covariant, only when summed up the gauge-covariance is restored. As a consequence, the Feynman diagrams that use free propagators of $\phi$ and add interaction with gauge field as a perturbation, don't give gauge-covariant results as well, unless summed up. This means that during the calculation of every single Feynman diagram, we needlessly calculate also some not-gauge-covariant expression that is bound to cancel out with other such expressions from other diagrams. In this work we are going to show that it is possible to present function $G_2(x_1,x_2)[A]$ in gauge-covariant form, and using that form to simplify some Feynman diagrams.

The method focuses on using the function $G(x_1,x_2)[A]$ in the form
\beq G_\phi(x_1,x_2)[A] = \Pexp\Big(\int_z^{x_1}\AAA\Big) \tilde G_\phi(z;x_1,x_2) \Pexp\Big(\int^z_{x_2}\AAA\Big) \label{GaugeCovForm}\eeq
where $\Pexp(\int\AAA)$ denotes a path-ordered exponential of the gauge field along a straight line, which can be defined by the following differential equation:
\beq \left\{\begin{array}{rl} \frac{\rd}{\rd\la}\Pexp\Big(\int_x^{x+\la a}\AAA\Big) &= a^\mu\AAA_\mu(x+\la a)\Pexp\Big(\int_x^{x+\la a}\AAA\Big)\\ \Pexp\Big(\int_x^x\AAA\Big) &= 1 \end{array}\right. \label{PathExpDef}\eeq
It can be shown that it satisfies
\beq \Pexp\Big(\int_{x_2}^{x_1}\AAA\Big) = U(x_1)\Pexp\Big(\int_{x_2}^{x_1}\AAA'\Big)U(x_2)^{-1} \label{PathExpWTIdentity2} \eeq
or
\begin{align} & \frac{\pd}{\pd x^\mu}\frac{\de \Pexp\Big(\int_{x_2}^{x_1}\AAA\Big)}{\de A^a_\mu(x)} - g f^{abc} A_\mu^b(x) \frac{\de \Pexp\Big(\int_{x_2}^{x_1}\AAA\Big)}{\de A^c_\mu(x)} = \nn\\
&\qquad = -\ri g \bigg(\Pexp\Big(\int_{x_2}^{x_1}\AAA\Big)T^a\de(x_2-x)-\de(x_1-x)T^a \Pexp\Big(\int_{x_2}^{x_1}\AAA\Big)\bigg) \label{PathExpWTIdentity} \end{align}
Combining equations (\ref{WTIdentity})--(\ref{PathExpWTIdentity2}) we get
\beq \tilde G(z;x_1,x_2)[A] = U(z)\tilde G(z;x_1,x_2)[A']U(z)^{-1} \eeq
\begin{gather}\begin{aligned} & \frac{\pd}{\pd x^\mu}\frac{\de \tilde G_\phi(z;x_1,x_2)[A]}{\de A^a_\mu(x)} - g f^{abc} A_\mu^b(x) \frac{\de \tilde G_\phi(z;x_1,x_2)[A]}{\de A^c_\mu(x)} = \\
&\qquad\qquad\qquad\qquad -\ri g \big[\tilde G_\phi(z;x_1,x_2)[A], T^a\big] \de(z-x) \label{GTildeWTIdentity} \end{aligned}\end{gather}
The simplification we have obtained with respect to eq. (\ref{WTIdentity}) gives us hope that $\tilde G_\phi(z;x_1,x_2)$ can be written in simpler form than $G(x_1,x_2)$. Especially if we focus on weak external fields $A$, and calculate $\tilde G_\phi(z;x_1,x_2)$ as a series in powers of $A(z)$ and its derivatives, we expect to see only gauge covariant structures, like $\AAA_{\mu\nu}$, $D_\mu \AAA_{\nu\rho}$ etc. \\
The point $z$ can be in principle arbitrary, but the easiest choice seems to be $z=x_1$, $z=x_2$ or $z=\frac{x_1+x_2}{2}$, though if we know beforehand in what Feynman diagram they will be needed, a better choice may be available. We will present the results for $z=\frac{x_1+x_2}{2}$, as in general, it is the one which simplifies many Feynman diagrams the most. In this case I'm going to write the function $\tilde G_\phi(\frac{x_1+x_2}{2};x_1,x_2)$ as a function of two arguments $\tilde G_\phi(\frac{x_1+x_2}{2},x_1-x_2)$, from now on.

To calculate $\tilde G_\phi(x,a)$, we move the path-ordered exponentials from (\ref{GaugeCovForm}) to the other side:
\beq \tilde G_\phi(x,a)[A] = \Pexp\Big(\int^x_{x+\frac{a}{2}}\AAA\Big) G_\phi(x+\frac12 a,x-\frac12 a) \Pexp\Big(\int_x^{x-\frac{a}{2}}\AAA\Big) \label{GphiEncased} \eeq
We will use $G_\phi(x+\frac12 a,x-\frac12 a)$ in the form of series in $A$, which comes naturally if we use Feynman diagrams. At the leading order we calculate only tree-level diagrams, with one scalar line and an arbitrary number of gauge field lines attached to it.
\begin{center}
\includegraphics{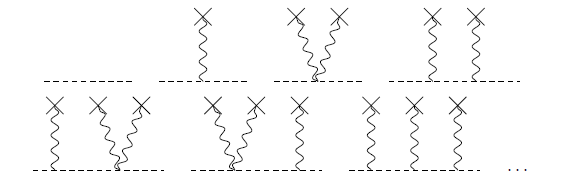} \hspace{.5cm} \dots\\
{\scriptsize {\it Fig. 1.} The tree-level diagrams that contribute to the scalar propagator in an external gauge field.}
\end{center} 
We also need to expand the formula in powers of momenta\slash derivatives of field $\AAA$. Let us remember that this is just $G(x_1,x_2)$, which doesn't yet transform simply under gauge transformations, and without path-ordered exponentials we don't expect to be able to gather different terms to create only gauge-covariant tensors like $\AAA_{\mu\nu}=\pd_\mu\AAA_\nu-\pd_\nu\AAA_\mu-[\AAA_\mu,\AAA_\nu]$. For now, let us just list all the terms that appear up to the order of $\cO(A^5)$, (counting derivatives and $A$ to be of the same order, as suggested by the form of the covariant derivative)\footnote{$\int_p = \int\frac{\rd^Dp}{(2\pi)^D}$}:
\begin{align}
& G_\phi(x+a\slash 2,x-a\slash 2)[A] = \nn\\
&= \ri\int_p \re^{-\ri p a} \bigg\{\frac{1}{p^2-m^2} + \frac{2\ri p_\mu}{(p^2-m^2)^2}\AAA^\mu(x) + \nn\\
&\qquad + \left(\frac{g_{\mu\nu}}{(p^2-m^2)^2}+\frac{-4p_\mu p_\nu}{(p^2-m^2)^3}\right)\AAA^\mu(x)\AAA^\nu(x) + \nn\\
&\qquad + \bigg(\frac{\ri g_{\al\be} p_\mu}{(p^2-m^2)^3}+\frac{-2\ri p_\al p_\be p_\mu}{(p^2-m^2)^4}\bigg) \pd^\al\pd^\be\AAA^\mu(x) + \nn\\
&\qquad + \bigg(\frac{2\ri p_\mu g_{\nu\al}}{(p^2-m^2)^3}+\frac{-4\ri p_\mu p_\nu p_\al}{(p^2-m^2)^4}\bigg) \big[\pd^\al\AAA^\mu(x),\AAA^\nu(x)\big] + \nn\\
&\qquad + \left(\frac{2\ri (p_\mu g_{\nu\rho}+p_\rho g_{\mu\nu})}{(p^2-m^2)^3} + \frac{-8\ri p_\mu p_\nu p_\rho}{(p^2-m^2)^4}\right)\AAA^\mu(x)\AAA^\nu(x)\AAA^\rho(x)+ \nn\\
&\qquad +  \bigg(\frac{\frac12 g_{\al\be} g_{\mu\nu}}{(p^2-m^2)^3} + \frac{-p_\al p_\be g_{\mu\nu}-p_\mu p_\al g_{\nu\be} - p_\mu  p_\be g_{\nu\al} - 3p_\mu  p_\nu g_{\al\be}}{(p^2-m^2)^4}+ \nn\\
&\qquad\qquad + \frac{8p_\mu  p_\nu p_\al p_\be}{(p^2-m^2)^5}\bigg)\big\{\pd^\al\pd^\be\AAA^\mu(x),\AAA^\nu(x)\big\} + \nn\\
&\qquad + \bigg(\frac{g_{\al\be} g_{\mu\nu}-g_{\nu\al}g_{\mu\be}}{(p^2-m^2)^3}+\frac{-2p_\al p_\be g_{\mu\nu}+2 p_\mu p_\be g_{\nu\al} + 2 p_\nu p_\al g_{\mu\be} -2 p_\mu p_\nu g_{\al\be}}{(p^2-m^2)^4} \bigg) \times \nn\\
&\qquad\qquad \times\big(\pd^\al\AAA^\mu(x)\big)\big(\pd^\be\AAA^\nu(x)\big) + \nn\\
&\qquad  + \bigg(\frac{g_{\rho\al} g_{\mu\nu}}{(p^2-m^2)^3} + \frac{-2 p_\rho p_\al g_{\mu\nu}-2 p_\mu p_\al g_{\nu\rho}-4p_\mu p_\nu g_{\rho\al}-4p_\mu p_\rho g_{\nu\al}}{(p^2-m^2)^4}+ \nn\\
&\qquad\qquad +\frac{16 p_\mu p_\nu p_\nu p_\al}{(p^2-m^2)^5}\bigg)\big(\pd^\al\AAA^\mu(x)\big)\AAA^\nu(x)\AAA^\rho(x) + \nn\\
&\qquad + \bigg(\frac{g_{\rho\al} g_{\mu\nu}- g_{\mu\al} g_{\nu\rho}}{(p^2-m^2)^3} + \frac{2 p_\mu p_\al g_{\nu\rho}-2 p_\rho p_\al g_{\mu\nu}-4p_\mu p_\nu g_{\rho\al}+4p_\nu p_\rho g_{\mu\al}}{(p^2-m^2)^4}\bigg)\times \nn\\
&\qquad\qquad \times\AAA^\mu(x)\big(\pd^\al\AAA^\nu(x)\big)\AAA^\rho(x) + \nn\\
&\qquad + \bigg(\frac{- g_{\mu\al} g_{\nu\rho}}{(p^2-m^2)^3} + \frac{2 p_\mu p_\al g_{\nu\rho}+2 p_\rho p_\al g_{\mu\nu}+4p_\mu p_\rho g_{\nu\al}+4p_\nu p_\rho g_{\mu\al}}{(p^2-m^2)^4} + \nn\\
&\qquad\qquad +\frac{-16 p_\mu p_\nu p_\nu p_\al}{(p^2-m^2)^5}\bigg)\AAA^\mu(x)\AAA^\nu(x)\pd^\al\AAA^\rho(x)  + \nn\\
&\qquad + \bigg(\frac{g_{\mu\nu}g_{\rho\si}}{(p^2-m^2)^3} + \frac{-4(p_\rho p_\si g_{\mu\nu}+p_\mu p_\si g_{\nu\rho}+p_\mu p_\nu g_{\rho\si})}{(p^2-m^2)^4} +\nn\\
&\qquad\qquad + \frac{16 p_\mu p_\nu p_\rho p_\si}{(p^2-m^2)^5}\bigg) \AAA^\mu(x)\AAA^\nu(x)\AAA^\rho(x)\AAA^\si(x)\bigg\} +\nn\\
&\qquad  + \cO\big((\pd,\AAA)^5\big) + (\text{boundary terms})\label{GSeries}
\end{align}
The boundary terms in this expression come from the ambiguity of redefining momenta in the derivation of this result; were all the integrals convergent, one could make the redefinition of integration variables like 
\begin{gather}\begin{aligned}
p_1 &\rarr p - \frac12 q - \la q \\
p_2 &\rarr p + \frac12 q + \la q \label{ShiftingMomenta}
\end{aligned}\end{gather}
for any value of $\la$, and obtain the same result. However, since some of the integrals are not convergent, we receive several (finite number)  terms of the form
\beq \la_i \int_p \frac{\pd}{\pd p_\mu} \Big(\re^{-\ri p a} \OOO^\mu(p,x)\Big) \eeq
where $\la_i$ are, at this point, arbitrary constants. We will get even more boundary terms in the formula for $\tilde G_\phi(x,a)$ because the path-ordered exponentials from (\ref{GphiEncased}) produce many terms that will need to be integrated by parts. For any operator $\OOO$ we have
\begin{gather}\begin{aligned}
& \Pexp\Big(\int^x_{x+\frac{a}{2}}\AAA\Big) \,\OOO\, \Pexp\Big(\int_x^{x-\frac{a}{2}}\AAA\Big) = \\
&= \OOO - \frac12 a_\mu \big\{\AAA^\mu(x),\OOO\big\} + \frac18 a_\mu a_\nu \Big(-\big[\pd^\mu\AAA^\nu(x),\OOO\big]+\big\{\AAA^\mu(x),\big\{\AAA^\nu(x),\OOO\big\}\big\}\Big) + \\
&\quad + \frac{1}{48}a_\mu a_\nu a_\rho\Big(-\big\{\pd^\mu\pd^\nu\AAA^\rho(x),\OOO\big\} +\big[\pd^\mu\AAA^\nu(x),\big\{\AAA^\rho(x),\OOO\big\}\big] + \\
&\quad\qquad +2\big\{\AAA^\mu(x),\big[\pd^\nu\AAA^\rho(x),\OOO\big]\big\} -\big\{\AAA^\mu(x),\big\{\AAA^\nu(x),\big\{\AAA^\rho,\OOO\big\}\big\}\big\}\Big) + \\
&\quad + \frac{1}{384}a_\mu a_\nu a_\rho a_\si \Big(-\big[\pd^\mu\pd^\nu\pd^\rho\AAA^\si(x),\OOO\big] + \big\{\pd^\mu\pd^\nu\AAA^\rho(x),\big\{\AAA^\si(x),\OOO\big\}\big\} + \\
&\quad\qquad +3\big\{\AAA^\mu(x),\big\{\pd^\nu\pd^\rho\AAA^\si(x),\OOO\big\}\big\} +3\big[\pd^\mu\AAA^\nu(x),\big[\pd^\rho\AAA^\si(x),\OOO\big]\big] + \\
&\quad\qquad - \big[\pd^\mu\AAA^\nu(x),\big\{\AAA^\rho(x),\big\{\AAA^\si,\OOO\big\}\big\}\big] -2\big\{\AAA^\mu(x),\big[\pd^\nu\AAA^\rho(x),\big\{\AAA^\si,\OOO\big\}\big]\big\} + \\
&\quad\qquad - 3\big\{\AAA^\mu(x),\big\{\AAA^\nu(x),\big[\pd^\rho\AAA^\si,\OOO\big]\big\}\big\} + \\
&\quad\qquad + \big\{\AAA^\mu(x),\big\{\AAA^\nu(x),\big\{\AAA^\rho,\big\{\AAA^\si,\OOO\big\}\big\}\big\}\big\}\Big) + \dots \label{2PathExpsSeries}
\end{aligned}\end{gather}
Then we use integration by parts to get rid of $a_\mu$ factors:
\beq a_\mu \int_p \re^{-\ri pa} f(p) = \int_p \re^{-\ri pa} \left(\frac{-\ri\pd}{\pd p^\mu}f(p)\right) + (\text{a boundary term}) \label{IntByParts}\eeq
As it turns out, with a general regularization\footnote{In regularizations that do not violate Ward-Takahashi identities, like the dimensional regularization, these terms will vanish and can be skipped early. In general case though they need to be remembered.} these boundary terms cannot be neglected and play a crucial role in some diagrams. They are also related to certain counterterms that appear in the process of renormalization, an example is presented in a section 4. below. For the list of boundary terms, see the appendix.

Eventually, many terms in the formula for $\tilde G_\phi(x,a)$ turn up to be vanishing, and the remaining ones can be grouped together to form gauge-covariant quantities, as expected:
\begin{gather}\begin{aligned}
\tilde G_\phi(x,a)[A] &= \ri\int_p \re^{-\ri p a} \bigg[\frac{1}{p^2-m^2} + D_\mu \AAA^{\mu\nu}(x) \frac{\frac23 \ri p_\nu}{(p^2-m^2)^3} + \\
&\qquad + \frac14\{\AAA_{\mu\nu}(x),\AAA_{\rho\si}(x)\}\left(\frac{g^{\mu\rho}g^{\nu\si}}{(p^2-m^2)^3}-\frac{4 g^{\mu\rho}p^\nu p^\si}{(p^2-m^2)^4}\right) + \\
&\qquad + (\text{terms of higher order in field $\AAA$ or its derivatives})\bigg] + \\
&\quad + (\text{boundary terms}) \label{ReducedScalarPropagator}
\end{aligned}\end{gather}
Let us behold how much simpler it has become, compared to (\ref{GSeries}). The reason for this is that there aren't many gauge-covariant structures that can be written at this order, and some of them (like $\AAA_{\mu\nu}$) cannot couple to a scalar because of the lack of proper Lorentz-invariant structures.

\section{Fermionic case}
The same thing can be done for fermions. We start with
\beq\cL_\psi[A] = \ri \overline\psi \ga^\mu D_\mu\psi - m\overline\psi\psi \label{LagrangianFermions}\eeq
\beq Z_\psi[\eta,\overline\eta,A] = \re^{\ri W_\psi[\eta,\overline\eta,A]} = \int \cD\psi\, \re^{\ri \int\rd^4x \big(\cL_\psi[A] + \overline\eta\psi + \overline\psi\eta\big)} \label{GeneratingFunctionalFermion} \eeq 
\beq G_\psi(x_1,x_2)[A] = -\ri\frac{\de^2W[\eta,\overline\eta,A]}{\de\overline\eta(x_1)\de\eta(x_2)}\bigg|_{\eta=0} \label{2PGFFermion} \eeq
Using the same method (we shall skip the intermediate steps), we can find that
\beq G_\psi(x+\frac{a}{2},x-\frac{a}{2})[A] = \Pexp\Big(\int_x^{x+\frac{a}{2}}\AAA\Big) \tilde G_\psi(x,a) \Pexp\Big(\int^x_{x-\frac{a}{2}}\AAA\Big) \label{GaugeCovFormFermion}\eeq
with
\begin{align}
& \tilde G_\psi(x,a) = \nn\\
&= \ri\int_p \re^{-\ri pa} \bigg\{\frac{\pslash+m}{p^2-m^2} - \frac12 \AAA_{\mu\nu}(x)\frac{\ga^{\mu\nu\al}p_\al+\ga^{\mu\nu}m}{(p^2-m^2)^2} + \nn\\
&\qquad +\frac{2\ri}{3} D_\mu\AAA_{\nu\rho}(x) \bigg[\Big(\frac{g^{\mu\rho}}{(p^2-m^2)^2}-\frac{p^\mu p^\rho}{(p^2-m^2)^3}\Big)\ga^\nu -\frac{g^{\mu\rho}p^\nu}{(p^2-m^2)^3}(\pslash+m) \bigg] + \nn\\
&\qquad + \frac18 D_\mu D_\nu\AAA_{\rho\si}(x) \bigg[\bigg(\frac{g^{\mu\nu}}{(p^2-m^2)^3}-\frac{4p^\mu p^\nu}{(p^2-m^2)^4}\bigg)(\ga^{\rho\si\al}p_\al+\ga^{\rho\si}m) + \nn\\
&\qquad\qquad - \frac{2p^\mu  p^\rho}{(p^2-m^2)^4}(\ga^{\nu\si\al}p_\al+\ga^{\nu\si}m) - \frac{2p^\mu  p^\rho}{(p^2-m^2)^4}(\ga^{\nu\si\al}p_\al+\ga^{\nu\si}m)  + \nn\\
&\qquad\qquad  - \frac{p^\mu\ga^{\nu\rho\si}+p^\nu\ga^{\mu\rho\si}}{(p^2-m^2)^3}\bigg]  + \nn\\
&\qquad + \big\{\AAA_{\mu\nu}(x),\AAA_{\rho\si}(x)\big\} \bigg[\frac{\ga^{\mu\nu\rho\si\al}p_\al+\ga^{\mu\nu\rho\si}m}{8(p^2-m^2)^3} + \frac{g^{\mu\rho}(p^\nu\ga^\si+p^\si\ga^\nu)}{2(p^2-m^2)^3} + \nn\\
&\qquad\qquad - \frac{g^{\mu\rho}p^\nu p^\si}{(p^2-m^2)^4}(\pslash+m)\bigg] + \nn\\
&\qquad + \big[\AAA_{\mu\nu}(x),\AAA_{\rho\si}(x)\big] \bigg(\frac{-\frac12 g^{\mu\rho}}{(p^2-m^2)^3}+\frac{p^\mu p^\rho}{(p^2-m^2)^4}\bigg)(\ga^{\nu\si\al}p_\al+\ga^{\nu\si}m) + \nn\\
&\qquad + (\text{terms of higher order in field $\AAA$ or its derivatives})\bigg\} + \nn\\
&\quad + (\text{boundary terms}) \label{ReducedFermionPropagator}
\end{align}
where $\ga^{\mu\nu}=\frac{1}{2!}\ga^{[\mu}\ga^{\nu]}$, $\ga^{\mu\nu\rho}=\frac{1}{3!}\ga^{[\mu}\ga^{\nu}\ga^{\rho]}$ etc. This formula is more complicated than (\ref{ReducedScalarPropagator}), because now all possible gauge-covariant structures appear, but still much simpler than the fermionic analog of (\ref{GSeries}) is. \\
We would like to stress that this formula is valid in any dimension, and it's independent of the definition of $\ga_5$. The only relation between gamma matrices that is necessary to derive it is their anticommutation relation, $\{\ga^\mu, \ga^\nu\}=2g^{\mu\nu}\one$. If the dimension is given, it can be simplified because of the fact that sufficiently long antisymmetrized products of gamma matrices are 0, for example, in 4 dimensions we have\footnote{with the convention that $\ga_5=\ri\ga^0\ga^1\ga^2\ga^3$, $\eps_{0123}=-\eps^{0123}=1$} $\ga^{\mu\nu\rho\si\al}=0$ and $\ga^{\mu\nu\rho\si} = \ri\eps^{\mu\nu\rho\si}\ga_5$.

\section{Boundary terms and counterterms}

As mentioned before, the boundary terms that appear in (\ref{ReducedScalarPropagator}) and (\ref{ReducedFermionPropagator}) cannot be neglected in a general case. Most of them are related to the counterterms necessary to make the effective theory of gauge field finite and/or gauge invariant. For the example, let us focus on the fermion theory in four dimensions. \\

From (\ref{GeneratingFunctionalFermion}), using explicit form of the lagrangian given by the (\ref{LagrangianFermions}), we can derive the equation
\beq \frac{\de W_\psi[\eta,\overline\eta,A]}{\de A_\mu^a(x)}\bigg|_{\eta=0} = g \Tr\Big\{\ga^\mu T^a \tilde G_\psi(x,a=0)\Big\} \label{DerWPsiOverA} \eeq
One of the boundary terms that appear in (\ref{ReducedFermionPropagator}) from integration by parts (\ref{IntByParts}) is
\beq \tilde G_\psi(x,a) \supset \AAA_\mu(x) \int_p \frac{\pd}{\pd p_\mu}\Big(\re^{-\ri p a} \frac{\pslash+m}{p^2-m^2}\Big) \label{BoundaryTermA} \eeq
We can see that this term will give a contribution to (\ref{DerWPsiOverA}):
\beq \frac{\de W_\psi[\eta,\overline\eta,A]}{\de A_\mu^a(x)}\bigg|_{\eta=0} \supset -2\ri g^2 A_\nu^a(x) \int_p \frac{\pd}{\pd p_\nu}\Big(\frac{p^\mu}{p^2-m^2}\Big) \eeq
which means that
\beq W_\psi[\eta,\overline\eta,A] \Big|_{\eta=0} \supset \frac12 \de M_A^2 \int_x A_\mu^a(x) A^{a\mu}(x) \label{MassTermForA}\eeq
with
\beq \de M_A^2 = -\ri\frac{g^2}{2}  \int_p \frac{\pd}{\pd p^\mu}\Big(\frac{p^\mu}{p^2-m^2}\Big) \label{deMA2General}\eeq
To maintain the gauge invariance of the effective theory of the gauge field, we need to add a counterterm to the lagrangian (\ref{LagrangianFermions}):
\begin{align} \cL_\psi &\rarr \cL_\psi + \cL_\text{ct} \nn \\
\cL_\text{ct} &\supset -\frac12 \de M_A^2\,\, A_\mu^a A^{a\mu} \label{MassACounterterm} \end{align}
It is exactly what we could get calculating the radiative correction to the gauge field mass from 1-loop diagram (Fig. 2.). However, the method presented in this paper shows that the value of the counterterm is already contained within the coefficient to a boundary term in fermion propagator calculated on the tree-level.
\begin{center}
\includegraphics{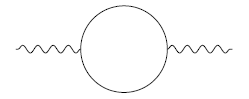} \\
{\scriptsize {\it Fig. 2.} A 1-loop diagram that contains a contribution to the gauge field mass term.}
\end{center}
The gauge symmetry will be preserved and there will be no need for such counterterm if the boundary terms vanish in a given regularization, for example in dimensional regularization. However, other regularizations can give non-zero results. 

\section{Boundary terms and axial anomaly}

In this section we are going to use the results obtained to show the origin of the ABJ-anomaly \cite{Adler, BellJackiw} in 4 dimensions. For simplicity's sake, we shall restrict ourselves to the case of abelian gauge group ($\AAA_{\mu\nu}=-\ri e F_{\mu\nu}$), but the calculations can be performed in a general case. With the propagator in the form (\ref{ReducedFermionPropagator}) it is easy to calculate
\begin{gather}\begin{aligned} & \langle J_5(x)\rangle = \langle \overline \psi(x)\ga_5\psi(x)\rangle = \\
&= -\Tr\Big\{\ga_5 G_\psi(x,x)\Big\} = -\Tr\Big\{\ga_5 \tilde G_\psi(x,a=0)\Big\} \label{J5} \end{aligned}\end{gather}
Up to the terms of order $\cO((\pd, A)^5)$, the only term from $\tilde G_\psi(x,a)$ that can contribute to this trace is
\beq \tilde G_\psi(x,a) \supset e^2\, m\, F_{\mu\nu}(x)F_{\rho\si}(x) \ga^{\mu\nu\rho\si} \int_p \re^{-\ri p a}\frac{-\ri }{4(p^2-m^2)^3} \eeq
Assuming that $\ga_5$ is defined\footnote{Which is how it should be defined in dimensional regularization, according to 't Hooft and Veltman \cite{tHooftVeltman}, where $\eps^{\mu\nu\rho\si}$ is defined in such a way that it s equal to 0 if any of its indices is different than 0, 1, 2 or 3.} such that
\beq \Tr\{\ga^{\mu\nu\rho\si}\ga_5\} = 4\ri\eps^{\mu\nu\rho\si}\eeq 
we find the result for (\ref{J5}) to be
\begin{gather}\begin{aligned} &\langle J_5(x)\rangle = \\
&= - e^2\, m\, F_{\mu\nu}(x)F_{\rho\si}(x)\, \Tr\{\ga_5\ga^{\mu\nu\rho\si}\} \cdot \frac{-1}{128\pi^2m^2} + \cO((\pd, A)^5) = \\
&=  \frac{\ri e^2}{32\pi^2m} \,\eps^{\mu\nu\rho\si} F_{\mu\nu}(x)F_{\rho\si}(x) + \cO((\pd, A)^5) \label{J5In4D}\end{aligned}\end{gather}

Independently, we can calculate
\begin{gather}\begin{aligned} & \langle J_5^\mu(x)\rangle = \langle \overline \psi(x)\ga^\mu\ga_5\psi(x)\rangle = \\
&= -\Tr\Big\{\ga^\mu\ga_5 G_\psi(x,x)\Big\} = -\Tr\Big\{\ga^\mu\ga_5 \tilde G_\psi(x,a=0)\Big\} \label{J5mu} \end{aligned}\end{gather}
This time the only terms that can contribute are some of the boundary terms  
\begin{gather}\begin{aligned} \tilde G_\psi(x,a) &\supset  e^2\, A_\mu(x) \pd_\nu A_\rho(x) \,\big(g^{\mu\al}\ga^{\nu\rho\be}+(\la_2-\la_3)g^{\nu\al}\ga^{\rho\mu\be}\big) \times \\
&\quad\qquad \times \int_p \frac{\pd}{\pd p^\al}\Big(\re^{-\ri pa}\frac{p_\be}{(p^2-m^2)^2}\Big) \end{aligned}\end{gather}
where $\la_2$ and $\la_3$ are arbitrary constants associated with the fact that shifting the integration variable (momentum $p$ in this case) in an integral that is not convergent produces a boundary term, as mentioned before (\ref{ShiftingMomenta}), see also the appendix. We get
\beq
\langle J_5^\mu(x)\rangle = (1+\la_2-\la_3) \ri e^2\,\eps^{\mu\nu\rho\si} A_\nu(x) \pd_\rho A_\si(x) \int_p \frac{\pd}{\pd p_\al}\Big(\frac{p_\al}{(p^2-m^2)^2}\Big) + \cO((\pd, A)^5) \label{J5muIn4D}\eeq
However $\langle J_5^\mu(x)\rangle$ needs to be gauge invariant, and to ensure that we must either put $\la_2-\la_3=-1$ or choose a regularization scheme in which this boundary term is equal to 0. Either way
\beq \langle J_5^\mu(x)\rangle = \cO((\pd, A)^5) \eeq

Let's now check the deviation from the naive axial current conservation equation:
\beq \langle \pd_\mu J_5^\mu(x)\rangle - 2\ri m \langle J_5(x)\rangle = 0 \label{AxialConservationNaive}\eeq
We can see it's not satisfied; instead
\beq \langle \pd_\mu J_5^\mu(x)\rangle - 2\ri m \langle J_5(x)\rangle = \frac{e^2}{16\pi^2} \,\eps^{\mu\nu\rho\si} F_{\mu\nu}(x)F_{\rho\si}(x) + \cO((\pd, A)^5) \label{AxialConservationAnomalous}\eeq
which is the well-known ABJ anomaly \cite{Adler}, \cite{BellJackiw}. The calculation here was made only up to the terms of order of $\cO((\pd, A)^5)$, but from the general theory \cite{PiguetSorella} we know there cannot be any terms of higher order in fields. The anomaly in this expression could be avoided but at the cost of losing the gauge invariance. It is important to understand that in this derivation the anomaly does not come from the divergence of the current, but from $\langle J_5(x)\rangle$, which is proportional to $1\slash m$. 

\section{Examples of the application of the method in Feynman diagrams}

Expressing the propagators in this form simplifies the calculation of certain Feynman diagrams. The most important limitation though is that we're using an expansion of $\tilde G(x,a)$ in the number of external gauge field lines and their momenta, so this method can be used only if the momenta of the gauge bosons are much smaller than the masses of particles they couple to. \\

A simple example would be the decay of a light neutral scalar coupled to heavy charged fermions:
\beq \cL = \frac12\pd^\mu\varphi\pd_\mu\varphi - \frac12 m_\varphi^2\varphi^2 + \ri\overline\psi\ga^\mu D_\mu\psi - m_\psi\overline\psi\psi + \cL_\text{int}\eeq
\beq \cL_\text{int} = \left\{\begin{array}{ll}-\la\varphi\overline\psi\psi&\text{for scalar}\\-\ri \la\varphi\overline\psi\ga_5\psi&\text{for pseudoscalar}\end{array}\right. \eeq
Normally, the calculation of the amplitude of $\varphi\rarr AA$ decay would require two triangle diagrams at 1-loop level (Fig. 3.), and neither of them is gauge-covariant (only their sum is), and the part cancelled by the counterterm is hidden within the expression. However, with the method we present in this work, we only need one tadpole diagram, which is gauge-covariant by itself, with the exception of contribution cancelled by a counterterm which is clearly visible (Fig. 4.). 
\begin{center}
\includegraphics{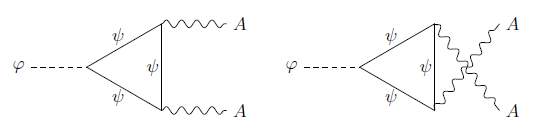} \\
{\scriptsize {\it Fig. 3.} Standard 1-loop diagrams that describe $\varphi\rarr AA$ decay. The propagators of $\psi$ are the free propagators in vacuum.} \\ \vspace{.2cm}
\includegraphics{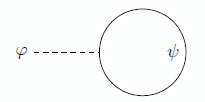} \\
{\scriptsize {\it Fig. 4.} The only 1-loop diagram that describe $\varphi\rarr AA$ decay with the method presented in this work. The propagator of $\psi$ is the propagator in external gauge field.} \\ 
\end{center} 

At the 1-loop level, in the case of the coupling without $\ga_5$, we have
\begin{align} &_\text{out}\langle AA|\varphi\rangle_\text{in} = \nn\\*
&= \ri \la \int_x \langle AA|\Tr\big\{G_\psi(x,x)\big\}|0\rangle \langle 0|\varphi(x)|\varphi\rangle = \nn\\
&= \ri \la \int_x \langle AA|\Tr\big\{\tilde G_\psi(x,0)\big\}|0\rangle \langle 0|\varphi(x)|\varphi\rangle = \nn\\
&= -8\ri \la \int_p \frac{g^{\mu\rho}p^\nu p^\si m_\psi }{(p^2-m_\psi^2)^4} \int_x \langle AA|\Tr\big\{\AAA_{\mu\nu}(x)\AAA_{\rho\si}(x)\big\}|0\rangle \langle 0|\varphi(x)|\varphi\rangle + \dots + \nn\\
&\quad + 2 \la \int_p \frac{\pd^2}{\pd p_\mu\pd p_\nu}\Big(\frac{m_\psi}{p^2-m_\psi^2}\Big) \int_x \langle AA|\Tr\big\{\AAA_\mu(x)\AAA_\nu(x)\big\}|0\rangle \langle 0|\varphi(x)|\varphi\rangle = \nn\\
&= -\frac{\la}{24\pi^2 m_\psi} \int_x \langle AA|\Tr\big\{\AAA^{\mu\nu}(x)\AAA_{\mu\nu}(x)\big\}|0\rangle \langle 0|\varphi(x)|\varphi\rangle + \dots + \\
&\quad + (\text{a part cancelled by a counterterm}) \nn \end{align}
where dots denote the terms with higher number of derivatives of $\AAA_\mu$. \\
The case of the coupling with $\ga_5$ is even simpler:
\begin{gather}\begin{aligned}& _\text{out}\langle AA|\varphi\rangle_\text{in} =\\
&= - \la \int_x \langle AA|\Tr\big\{\ga_5 G_\psi(x,x)\big\}|0\rangle \langle 0|\varphi(x)|\varphi\rangle = \\
&= - \la \int_x \langle AA|\Tr\big\{\ga_5 \tilde G_\psi(x,0)\big\}|0\rangle \langle 0|\varphi(x)|\varphi\rangle = \\
&= \frac14 \la \Tr\big\{\ga_5\ga^{\mu\nu\rho\si}\big\}\int_p \frac{m_\psi }{(p^2-m_\psi^2)^3} \times \\
&\qquad\qquad\qquad \times\int_x \langle AA|\Tr\big\{\AAA_{\mu\nu}(x)\AAA_{\rho\si}(x)\big\}|0\rangle \langle 0|\varphi(x)|\varphi\rangle + \dots = \\
&=  \frac{\la}{16\pi^2 m_\psi} \int_x \langle AA|\Tr\big\{\AAA^{\mu\nu}(x)\tilde\AAA_{\mu\nu}(x)\big\}|0\rangle \langle 0|\varphi(x)|\varphi\rangle + \dots \end{aligned}\end{gather} 

Another calculation in this model that becomes much simpler if we use formulae derived in this work, is the correction to the mass of $\varphi$ due to background gauge field that appears in the effective action after integrating out the $\psi$ field. With this method, it is all contained within a single diagram (Fig. 5.)
\begin{center}
\includegraphics{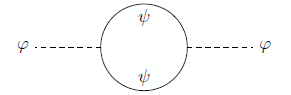} \\
{\scriptsize {\it Fig. 5.} A 1-loop diagram that contains the corrections to the mass term of $\varphi$.}
\end{center}
The contribution to the effective action given by this diagram is (assuming the coupling without $\ga_5$):
\begin{gather}\begin{aligned}
\cS_\text{eff} &\supset -\frac12 \ri \la^2 \int_x \int_y \varphi(x)\varphi(y) \Tr\big\{G_\psi(x,y)G_\psi(y,x)\big\} = \\
&= -\frac12 \ri \la^2 \int_x \int_a \varphi(x+a\slash 2)\varphi(x-a\slash 2) \Tr\big\{\tilde G_\psi(x,a)\tilde G_\psi(x,-a)\big\} = \\
&= -\frac12 \ri \la^2 \int_x \varphi(x)^2 \int_a \Tr\big\{\tilde G_\psi(x,a)\tilde G_\psi(x,-a)\big\} + \dots
\end{aligned}\end{gather}
For $a\neq 0$ the boundary terms disappear because of the oscillating factor $\re^{-\ri pa}$, so the only relevant terms that contribute to this integral are covariant terms from (\ref{ReducedFermionPropagator}). The first non-zero terms after the field-independent term that we get from standard Feynman diagram are proportional to $\AAA_{\mu\nu}(x)\AAA_{\rho\si}(x)$:
\begin{gather}\begin{aligned}
& \int_a \Tr\big\{\tilde G_\psi(x,a)\tilde G_\psi(x,-a)\big\} = \\
&= \int_p \bigg(\Tr\Big\{\Big(\ri\frac{\pslash+m_\psi}{p^2-m_\psi^2}\Big)^2\Big\} + \Tr\Big\{\Big(\frac{-\ri}{2}\AAA_{\mu\nu}(x)\frac{\ga^{\mu\nu\al}p_\al+\ga^{\mu\nu}m_\psi}{(p^2-m_\psi^2)^2}\Big)^2\Big\} + \\
&\quad + 2\Tr\Big\{\ri\frac{\pslash+m_\psi}{p^2-m_\psi^2} \times \ri\{\AAA_{\mu\nu}(x),\AAA_{\rho\si}(x)\}\Big(\frac{g^{\mu\rho}(p^\nu\ga^\si+p^\si\ga^\nu)}{2(p^2-m_\psi^2)^3} + \\
&\qquad\qquad\qquad\qquad\qquad\qquad - \frac{g^{\mu\rho}p^\nu p^\si(\pslash+m_\psi)}{(p^2-m_\psi^2)^4}\Big)\Big\} + \dots\bigg) = \\
&= (\text{a constant that is a subject to renormalization}) + \\
&\quad + \frac{\ri}{24\pi^2m_\psi^2} \Tr\{\AAA^{\mu\nu}(x)\AAA_{\mu\nu}(x)\} + \dots 
\end{aligned}\end{gather}
Therefore the renormalized contribution to the effective action is:
\beq\cS_\text{eff} \supset \int_x  \frac{\la^2 }{48\pi^2m_\psi^2} \varphi(x)^2 \Tr\{\AAA^{\mu\nu}(x)\AAA_{\mu\nu}(x)\} + \dots
\eeq
With the propagator already in the form (\ref{ReducedFermionPropagator}) the computation is much simpler than it would be to do from the scratch.

\section{Complex diagrams}

In the diagrams in the previous section, in which there were no more than two vertices (not counting the interaction with the gauge field), and none of them contained the gauge group generators, the formulae drastically simplified because the path exponents cancel each other and disappear from the expression. In more complicated diagrams they may remain but even then there is a way to combine them to quasi-local, gauge covariant expressions if necessary.
\begin{center}
\includegraphics{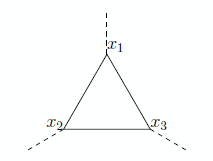} \\
{\scriptsize {\it Fig. 6.} The simplest case in which we need to remember about Wilson lines.}
\end{center}
If we have more than two vertices, like in the Fig. 6., the first thing that we may notice is that their middle points do not coincide, for example one of them can be
\beq \Pexp\Big(\int^{x_1}_{\frac{x_1+x_2}{2}}\AAA\Big)\tilde G(\frac{x_1+x_2}{2},x_1-x_2)\Pexp\Big(\int^{\frac{x_1+x_2}{2}}_{x_2}\AAA\Big) \label{Propx1x2} \eeq and the other
\beq \Pexp\Big(\int^{x_2}_{\frac{x_2+x_3}{2}}\AAA\Big)\tilde G(\frac{x_2+x_3}{2},x_2-x_3)\Pexp\Big(\int^{\frac{x_2+x_3}{2}}_{x_3}\AAA\Big) \label{Propx2x3} \eeq
If $x_1\neq x_3$ then one of them is expressed in terms of the gauge field in point $\frac{x_1+x_2}{2}$, and the other in point $\frac{x_2+x_3}{2}$. To solve this issue, we can use the analog of Taylor series in the space with a gauge connection:
\begin{gather}\begin{aligned} \cO(x) &= \Pexp\Big(\int^{x}_{x_0}\AAA\Big)\times \bigg(\cO(x_0) + (x-x_0)^\mu D_\mu\cO(x_0) + \\
&\qquad\qquad + \frac12 (x-x_0)^\mu (x-x_0)^\nu D_\mu D_\nu \cO(x_0)+\dots\bigg) \times \Pexp\Big(\int^{x_0}_{x}\AAA\Big) \label{ShiftOperator} \end{aligned}\end{gather}
We can choose the point $x_0$ arbitrarily, usually one of the vertices or the diagram's "mass center" is a choice that leads to simple expressions later. We can also use this formula to "shift" any vertex that contain group indices to the point $x_0$.\\
This way, if the initial formula contains expressions like
\beq \dots \cO_{12}(\frac{x_1+x_2}{2}) \Pexp\Big(\int^{\frac{x_1+x_2}{2}}_{x_2}\AAA\Big) V_2(x_2) \Pexp\Big(\int^{x_2}_{\frac{x_2+x_3}{2}}\AAA\Big) \cO_{23}(\frac{x_2+x_3}{2}) \dots \eeq
they can be written in the following form:
\begin{gather}\begin{aligned} &\dots \cO'_{12}(x_0) \times \Pexp\Big(\int^{x_0}_{\frac{x_1+x_2}{2}}\AAA\Big) \Pexp\Big(\int^{\frac{x_1+x_2}{2}}_{x_2}\AAA\Big) \Pexp\Big(\int^{x_2}_{x_0}\AAA\Big) \times \\
&\times V'_2(x_0) \times \Pexp\Big(\int^{x_0}_{x_2}\AAA\Big) \Pexp\Big(\int^{x_2}_{\frac{x_2+x_3}{2}}\AAA\Big) \Pexp\Big(\int^{\frac{x_2+x_3}{2}}_{x_0}\AAA\Big) \times \cO'_{23}(x_0) \dots\end{aligned}\end{gather}
with $\cO'$ and $V'$ derived from $\cO$ and $V$ as eq. (\ref{ShiftOperator}) dictates.
As we can see, the path-ordered exponentials form triangles. These triangles can be expressed in the form of a quasi-local, gauge-covariant series at the point $x_0$:
\begin{align} & \Pexp\Big(\int^{x_0}_{x_1}\AAA\Big) \Pexp\Big(\int^{x_1}_{x_2}\AAA\Big) \Pexp\Big(\int^{x_2}_{x_0}\AAA\Big) = \nn \\
&= 1 - \frac14\Big((x_1-x_0)^\mu(x_2-x_0)^\nu - (x_2-x_0)^\mu(x_1-x_0)^\nu\Big) \AAA_{\mu\nu}(x_0) + \nn \\
&\quad -\frac{1}{12} (x_1+x_2-2x_0)^\mu\Big((x_1-x_0)^\nu(x_2-x_0)^\rho - (x_2-x_0)^\nu(x_1-x_0)^\rho\Big)\times \nn\\
&\quad\qquad \times D_\mu\AAA_{\nu\rho}(x_0) + \label{OpenWilsonLoop} \\
&\quad + \dots \nn \end{align}
Another approach, that would let us avoid such triangles, would be to use a general formula (\ref{GaugeCovForm}) with point $z=x_0$ being the same for all propagators in the diagram, instead of (\ref{Propx1x2}) and (\ref{Propx2x3}). Then only the vertices need to be shifted with (\ref{ShiftOperator}), and the necessary path-ordered exponentials are already there. However, general formula for $\tilde G(x_0;x_1,x_2)$ is more complicated than in the case of $x_0=\frac{x_1+x_2}{2}$ and contains vectors $(x_0-x_1)^\mu$ and $(x_0-x_2)^\mu$. \\
\indent Either way, we obtain the formula in the form of a series in the powers of the field $\AAA_{\mu\nu}(x_0)$ and its derivatives. The formula however also contains a number of vectors $(x_i-x_j)^\mu$. If we want to get rid of them to perform the integrations over $x_i$ and remain with a single spatial integral, we can chose $x_0$ to be some linear combination of $x_1,\dots x_n$. Then all these vectors can first be decomposed into the vectors related to particular propagators, and then be turned into the derivatives over momenta using eq. (\ref{IntByParts}). \\
\indent While this procedure certainly looks complicated, it needs to be said that it is necessary only in the case of very complicated diagrams, with at least 3 vertices other than coupling to the external gauge field. And even then, if we are interested only in the leading contribution for weak field $\AAA_{\mu\nu}$, or we are working with the abelian case, there is a good chance we would be able to perform additional simplifications and reduce the number of the triangles of path-ordered exponentials before employing the formula (\ref{OpenWilsonLoop}). We should also remember that calculating such diagrams in the standard way is usually even more complicated, and one needs to calculate more diagrams to get a gauge-covariant result. With this method we work with gauge-covariant quantities all long, and we avoid computing some irrelevant contribution that cancel between standard diagrams at the end.

\section{Summary}

In this work i present an alternative method of calculating some amplitudes in QFT. While it does not make it possible to calculate anything that couldn't be calculated before, it can make some calculations faster, and most importantly, we can have better control on what happens in intermediate stages of computations, since the physically relevant, gauge-invariant terms are clearly visible and not masked by the multiple other terms cancelling in the final result. I only calculate what is going to remain, and do not need to consider irrelevant terms. Without the need to calculate all the diagrams to get a gauge-covariant results, it may also be easier to focus only on some subset of them, in situations when we are able to argue that others do not produce the contributions that are relevant for the case in hand or that they give contributions that are negligible. \\
\indent The method has its limitations of course. Because I only managed to derive the Feynman propagators in the form of a series in gauge fields and its deivatives, it's only apllicable when the external gauge fields is slowly changing, or the momenta of the external gauge bosons are much smaller than the masses of particles they couple to. For this reason, some processes, like the production of gauge bosons in high-energy collisions, can't be calculated with this method unless a better formula for the gauge-field-dependent propagators were to be derived. Still, it can be used in many other situations and I believe the clarity and simplifications brought by the application of the propagators in the form presented in this work makes the effort of deriving them worthwhile. 

\vspace{1cm}
\noindent{\bf {Acknowledgements:}}
I would like to thank Adrian Lewandowski, Krzysztof A. Meissner and Hermann Nicolai for the discussions and the help in preparing this paper.

\appendix

\section{Boundary terms}

For the purpose of reducing the size of the following expression, We shall denote
\beq E = \re^{-\ri p a}, \qquad D = \frac{1}{p^2-m^2}, \qquad S = \frac{\pslash+m}{p^2-m^2}, \qquad S^{\mu\nu} = S\ga^{\mu}S\ga^{\nu}S \eeq

We have 
\begin{align}
\tilde G_\phi(x,a) &= (\text{gauge covariant part, see eq. (\ref{ReducedScalarPropagator})}) + \nn\\
&\quad + \AAA_\mu(x) \int_p \frac{\pd}{\pd p_\mu}\big(E D\big) + \nn\\
&\quad - \ri \la_1 \pd_\mu\AAA_\nu(x) \int_p \frac{\pd}{\pd p_\mu}\Big(E \frac{\pd D}{\pd p_\nu}\Big) + \nn\\
&\quad - \frac{\ri}{4} \big\{\AAA_\mu(x),\AAA_\nu(x)\big\} \int_p \frac{\pd^2}{\pd p_\mu\pd p_\nu}\big(E D\big) + \dots
\end{align}

\begin{align}
\tilde G_\psi(x,a) &= (\text{gauge covariant part, see eq. (\ref{ReducedFermionPropagator})}) + \nn\\
&\quad + \AAA_\mu(x) \int_p \frac{\pd}{\pd p_\mu}\big(E S\big) + \nn\\
&\quad - \ri \la_1 \pd_\mu \AAA_\nu(x) \int_p \frac{\pd}{\pd p_\mu}\Big(E \frac{\pd S}{\pd p_\nu}\Big) + \nn\\
&\quad - \frac{\ri}{4} \big\{\AAA_\mu(x),\AAA_\nu(x)\big\} \int_p \frac{\pd^2}{\pd p_\mu\pd p_\nu}\big(E S\big) + \nn\\
&\quad - \frac{1}{24} \pd_\mu\pd_\nu\AAA_\rho(x) \int_p \bigg(\frac{\pd^3 E}{\pd p_\mu\pd p_\nu\pd p_\rho} S + E\frac{\pd^3 S}{\pd p_\mu\pd p_\nu\pd p_\rho} \bigg) + \nn\\
&\quad + \frac14 \big\{\pd_\mu\AAA_\nu(x),\AAA_\rho(x)\big\} \int_p \frac{\pd}{\pd p_\rho}\Big(E S^{[\mu\nu]}\Big) + \nn\\
&\quad + \big[\pd_\mu\AAA_\nu(x),\AAA_\rho(x)\big] \int_p \bigg(\frac{1}{24}\frac{\pd^3 E}{\pd p_\mu\pd p_\nu\pd p_\rho} S +\frac18\frac{\pd^2 E}{\pd p_\mu\pd p_\nu} \frac{\pd S}{\pd p_\rho} + \nn\\
&\quad\qquad - \frac{1}{12} E\frac{\pd^3 S}{\pd p_\mu\pd p_\nu\pd p_\rho} \bigg) + \nn\\
&\quad + \la_2 \pd_\mu \AAA_\nu(x) \AAA_\rho(x) \int_p \frac{\pd}{\pd p_\mu}\Big(E S^{\nu\rho}\Big) + \nn\\
&\quad + \la_3 \AAA_\rho(x) \pd_\mu \AAA_\nu(x) \int_p \frac{\pd}{\pd p_\mu}\Big(E S^{\rho\nu}\Big) + \nn\\
&\quad + \AAA_\mu(x)\AAA_\nu(x)\AAA_\rho(x) \int_p\bigg(-\frac16\frac{\pd^3 E}{\pd p_\mu\pd p_\nu\pd p_\rho} S - \frac18\frac{\pd^2 E}{\pd p_\mu\pd p_\nu} \frac{\pd S}{\pd p_\rho} + \nn\\
&\quad\qquad -\frac14 \frac{\pd^2 E}{\pd p_\mu\pd p_\rho} \frac{\pd S}{\pd p_\nu} - \frac18 \frac{\pd^2 E}{\pd p_\nu\pd p_\rho} \frac{\pd S}{\pd p_\mu} + \frac13 E\frac{\pd^3 S}{\pd p_\mu\pd p_\nu\pd p_\rho\nn}\\
&\quad\qquad -\frac12 \frac{\pd}{\pd p_\mu} \Big(E\,S^{\nu\rho}\Big) -\frac12 \frac{\pd}{\pd p_\rho} \Big(E\,S^{\mu\nu}\Big)\bigg) + \dots
\end{align}
While not all of these formulae look explicitly like boundary terms, performing the integrations by parts shows that they are indeed. The boundary terms that contain higher number of fields $\AAA$ and their derivatives can be omitted in 4 dimensions, as they are either boundary terms of convergent integrals or vanish for $a=0$.

\end{document}